\documentstyle[12pt]{article}
\begin{document}

\begin{center}
{\Large {\bf Thermodynamic properties of $d_{x^{2}-y^{2}}+id_{xy}$ 
Superconductor}}\\ 
\vspace{1.0cm} 
{\bf Tulika Maitra}\footnote{email: tulika@phy.iitkgp.ernet.in}\\ 

\noindent Department of Physics \& Meteorology \\  
Indian Institute of Technology,
Kharagpur 721302 India \\   

\end{center}  
\begin{abstract}
In view of the current interest in $d_{x^{2}-y^{2}}+id_{xy}$ superconductors
some of their thermodynamic properties have been studied to obtain relevant
information for experimental verification. The temperature 
dependence of the specific heat and superfluid density show marked 
differences in $d_{x^{2}-y^{2}}+id_{xy}$ state compared to the pure d-wave 
state.  A second order phase transition is observed on lowering the temperature
into a $d_{x^{2}-y^{2}}+id_{xy}$ state from the $d_{x^{2}-y^{2}}$ state 
with the opening up of a gap all over the fermi surface. The  
thermodynamic quantities in $d_{x^{2}-y^{2}}+id_{xy}$ state are dominated
by this gap as in an s-wave superconductor as opposed to the algebraic
temperature dependence in pure d-wave states coming from the low energy 
excitations across the node(s). 

\end{abstract}  
\vspace{.5cm} 

\noindent {\bf Introduction}  
\vspace{.5cm} 

The series of experiments carried out over the last few years to establish
the nature of symmetry in high temperature superconductors with their level
of sophistication and ingenuity have thrown up new challenges towards an
understanding of the physics of these systems\cite{ann,vanh}. New findings 
have 
come up  with surprising regularity with the latest one being the observation
of a plateau in the thermal conductivity by Krishana et al.\cite{kris}
and its subsequent interpretation in terms of the appearance of a time-reversal
symmetry breaking state ($d_{x^{2}-y^{2}}+id_{xy}$)\cite{laugh}.  

An understanding of the pairing mechanism that underlies the superconducting
instability is essential for the emergence of a microscopic 
theory for these superconductors. Countless experiments have been performed 
and various theoretical models have been proposed to probe the symmetry of 
the OP in these highly anisotropic 
unconventional superconductors. At present it is almost universally accepted
that the OP is highly anisotropic with a symmetry of the d-wave\cite{ann}. 

In the recent experiment of Krishana et. al.\cite{kris} thermal 
conductivity 
as functions of both magnetic field and temperature has been measured on 
a sample of high $T_c$ superconducting material $Bi_2Sr_2CaCu_2O_8$. 
They observed that the thermal conductivity initially decreases with the 
increase of magnetic field and above a particular value of the field, which
depends on temperature, thermal conductivity becomes independent of 
field. These observations gave an indication that the material 
undergoes a phase transition in presence of the magnetic field. The authors 
suggested that this magnetically induced phase might have a complex order 
parameter symmetry
such as $d_{x^{2}-y^{2}}+id_{xy}$ or $d_{x^{2}-y^{2}}+is$ where the gap is 
nonzero on the entire FS. Corroboration of these results came quickly from 
other groups as well\cite{aubi}.

Laughlin\cite{laugh} showed that in 
presence of magnetic field the new superconducting phase must have an OP 
that violates both time reversal and parity and is of $d_{x^{2}-y^{2}}+id_{xy}$
 symmetry.
There were earlier predictions for such a state in the region near a grain 
boundary where the gap has sharp variation across it\cite{sigrist} with a
spontaneous current generated along the boundary and in a doped Mott insulator
with short range antiferromagnetic spin correlation\cite{laugh2,rokh}.

It would, therefore, be interesting to study the thermodynamic behaviour of such
superconductors having OP symmetry of $d_{x^2-y^2}+id_{xy}$ type to provide
further experimental observations to confirm its existence.
We use the usual weak coupling theory to obtain the gap functions in the
region of parameter space where a $d+id$ state is a stable one, and calculate
thermodynamic quantities like the specific heat and superfluid density 
(and hence the penetration depth) and contrast them with a pure d-wave state.     

\vspace{.5cm} 
\noindent {\bf Model and Calculations} 
\vspace{.5cm} 

Unlike pure d-wave OP, the $d_{x^{2}-y^{2}}+id_{xy}$ gap function has no 
node along the FS.
The OP has non-zero magnitude all over but it changes sign in each quadrant 
of the Brillouin zone, while a pure s-wave OP does also have  non vanishing 
magnitude but its sign remains same throughout.
This non vanishing gap inhibits creation of quasiparticle excitations
at low energies whereas in a 
pure d-wave state the gapless excitations are available in large 
numbers at low  energies due to the presence of line nodes.
Taking a tight binding model for a 2-dimensional square lattice,   various 
physical quantities have been calculated within the framework of the usual
weak coupling theory. The effective interaction has been taken 
in the separable form\cite{kot} and expanded in the relevant basis functions
of the irreducible representation of $C_{4v}$.  
$$V{(\bf k-k')}=\sum_{i=1,2}V_{i}\eta_{i}{(\bf k)}\eta_{i}{(\bf k')}$$ 
\noindent where 
$\eta_{1}{(\bf k)}=\frac{1}{2}(cosk_{x}-cosk_{y})$ and 
$\eta_{2}{(\bf k)}=sink_{x}sink_{y}$ (respectively for $d_{x^{2}-y^{2}}$ and 
$d_{xy}$ symmetries). $V_{1}/8$ and
$V_{2}/8$ are the respective coupling strengths for the near-neighbour and
next near-neighbour interactions.
The coupling strengths have been chosen in such a way as to allow both 
the components of the OP to exist simultaneously and the superconducting
transition temperature of $d_{xy}$ component to be lower than that of 
$d_{x^2-y^2}$ and is in the range of the observed values.
Considering only the nearest neighbour hopping, the band dispersion  
is $\epsilon_{\bf k}=-2t(cosk_{x}+cosk_{y})$ where $t$ is the nearest 
neighbour hopping integral and expanding the OP as 
$\Delta_{\bf k}=\sum_{i=1,2}\Delta_{i}\eta_{i}{(\bf k)}$ for $\eta_{i}({\bf k})$
defined above (for the $d_{x^{2}-y^{2}}+id_{xy}$ symmetry), the 
standard mean-field gap equation becomes a set of two coupled equations
$$\Delta_{1}=-V_{1} {\bf \sum_k} \frac{\Delta_{1}}{2E_{\bf k}} 
\eta_{1}^2({\bf k})tanh\left(\frac{E_{\bf k}}{2k_BT}\right)$$
\noindent and $$\Delta_{2}=-V_{2}{\bf \sum_k}\frac{\Delta_{2}}{2E_{\bf k}}
\eta_{2}^2({\bf k})tanh\left(\frac{E_{\bf k}}{2k_BT}\right).$$ 
\noindent Here the quasiparticle spectrum in the ordered state is given by  
$E_{\bf k}=\sqrt{{(\epsilon_{\bf k}-\mu)}^{2}+|\Delta_{k}|^{2}}$ 
where $\mu$ is the chemical potential. The coupled set of gap equations are 
solved numerically in a selfconsistent manner 
with the parameters $t=0.15$ eV, $V_{1}=0.445t$ eV and $V_{2}=3.202t$ eV.
The solutions give the expected square root temperature dependences of the 
two components of the order parameter $d_{x^2-y^2}$($\Delta_{1}$) and 
$d_{xy}$($\Delta_{2}$) and the corresponding $T_c$s ($T_{c1}$ and $T_{c2}$) 
as shown in Fig. 1. It is to be noted that the consistent solutions exist
for both $d_{x^2-y^2}$ and $d_{xy}$ components of the OP for a very narrow
range of $V_1$ and $V_2$. The $d_{xy}$ component of the OP exists
only when the next nearest neighbour interaction ($V_2$) is taken into account.
It has also been observed that the solutions have sensitive dependence on the 
values of the chemical potential and the next nearest neighbour 
hopping integral($t'$). To be more specific, if we change the value of chemical 
potential to $-0.25$ eV from $\mu = 0$ with $t'=0$, the $d+id$ state ceases 
to exist, but the inclusion of the $t'$ term (with $t'=0.4t$) brings 
the $d+id$ state back.  The solutions, of course, exist only for a narrow 
range of values of 
the chemical potential: for instance $\mu=-0.22$ eV to $\mu=-0.26$ eV with
$t'=0.4t$ has well defined solutions with $\Delta_1(0)>\Delta_2(0)$. 
Similarly if we keep the value of $\mu$ fixed at any of the above 
values and start changing the value of $t'$, only a very narrow range of
$t'$ gives us a $d+id$ solution. This interplay of $t'$ and $\mu$ is dictated
by the location of the van Hove singularity (vHS) with respect to the 
fermi energy.

The quasiparticle spectrum along different directions in 
the first quadrant of the Brillouin zone is shown in Fig. 2 and the finite
gap along all $k-$points is clearly visible.
With the excitation spectrum thus obtained, it is straightforward to calculate
thermodynamic quantities, namely, the specific heat and superfluid
density, in the different ordered states. From the usual definition 
in terms of the 
derivative of entropy\cite{tin}, we calculate the specific heat across the 
transitions and show it in Fig. 3. Two sharp jumps in the specific heat curve 
are observed at the respective transition temperatures.

The superfluid density $\rho_{s}$(T) has been calculated using the 
standard techniques of many body theory\cite{scal,tar}. In the presence of a
transverse vector potential with the chosen gauge 
$A_{y}=0$, the hopping matrix element($t_{ij}$) for the kinetic energy term in
the Hamiltonian$(H_{0})$ is modified by the Peierl's phase 
factor $exp[\frac{ie}{\hbar c}\int_{{\bf r}_{j}}^{{\bf r}_{i}}{\bf A}.d{\bf l}]
$. The total current (in the linear response) $J_{x}({\bf r_{i}})$ produced by
the potential consists of both the diamagnetic and paramagnetic terms and 
can be derived by differentiating $H_{0}$ with respect to $A_{x}({\bf r}_{i})$. 
Hence
$$j_{x}({\bf r}_{i})=-c {\frac{\partial{H_{0}}}{\partial{A_{x}}({\bf r}_{i})}}
={j_{x}}^{para}({\bf r}_{i})+{j_{x}}^{dia}({\bf r}_{i})$$ 
\noindent where the paramagnetic current in the long 
wavelength limit in the linear response is given by
$${\bf j}_{x}^{para}({\bf q})=-\frac{i}{c}\,\,lim_{q\rightarrow 0}
lim_{\omega\rightarrow 0}\int d\tau\theta(\tau)e^{i\omega\tau} \langle 
[j_{x}^{para}({\bf q},\tau),j_{x}^{para}(-{\bf q},0)]\rangle 
{\bf A_{x} ({\bf q})},$$ 
\noindent and the diamagnetic part is given by
$${\bf j}_{x}^{dia}({\bf q})= -\frac{e^{2}}{N\hbar^{2}c}\sum_{{\bf k},\sigma}
\langle c^{\dagger}_{{\bf k},\sigma} c_{{\bf k},\sigma}\rangle{\frac{\partial^{2}\epsilon_{\bf k}}{\partial^{2}{k_{x}}^{2}}}{\bf A_{x}({\bf q})}.$$
  
\noindent Here the averaging is done in the mean-field superconducting state.
Fig. 4 shows the variation of $\rho_{s}$ with temperature. At low temperatures 
where the superconductor is in $d_{x^2-y^2}+id_{xy}$ state, the superfluid 
density 
exhibits an exponential decay reflecting the gapped excitations. Above the 
second transition temperature($T_{c2}$) 
at which the $d_{xy}$ component of the OP vanishes and the 
superconductor undergoes a transition to $d_{x^2-y^2}$ phase, the superfluid 
density curve shows a power law behaviour expected from the low energy 
quasiparticles.
\vspace{.5cm} 

\noindent {\bf Results and Discussion} 
\vspace{.5cm} 

The self-consistent solutions for the order parameters (Fig. 1) show that as we
decrease the temperature, first there is a continuous transition into a
superconducting state where the OP is of $d_{x^2-y^2}$ symmetry with no
$d_{xy}$ component. On further decreasing the temperature a second continuous
transition occurs and the $d_{xy}$ component appears (with a phase $\pi/2$
with respect to the $d_{x^2-y^2}$ component) breaking the time reversal 
symmetry.
A stable $d+id$ phase does not exist unless the next nearest
neighbour interaction is being considered. This is because the next nearest
neighbour attraction accounts for the pairing along the (110) direction.
The sensitive dependence of the solutions on the chemical potential and the
next near neighbour hopping integral is understood by studying the nature of
the non-interacting density of states (DOS).
It has been noticed that the van Hove singularity(vHS) in the non-interacting
DOS lies far away from the fermi level when we include the $t'$ term in the 
band keeping the chemical potential zero, but if in addition we change the
the chemical potential to $-0.25$ eV, the vHS moves close to 
the fermi level. 

In the $d_{x^{2}-y^{2}}+id_{xy}$ state there exists no node on the FS, a  
gap opens throughout. Hence the low energy quasiparticle 
excitations are exponentially down in comparison to the pure d-wave state that
has line nodes on the FS. This is borne out from the plot of the quasiparticle 
energy spectrum along different directions of BZ (Fig. 2).

As temperature decreases from $T_c$ corresponding to the $d_{x^2-y^2}$ state, 
the low energy quasiparticle excitations are exponentially low in the 
$d_{x^{2}-y^{2}}+id_{xy}$ state due to the appearance of an additional OP 
of $d_{xy}$ symmetry and phased by $90$ degree with the existing 
$d_{x^{2}-y^{2}}$ OP. The thermodynamic quantities are therefore affected in 
this new state quite severely. The temperature dependence of the specific heat 
(Fig. 3) shows the difference. The sharp jumps at transition temperatures in 
the specific heat curve, are clear indication of second order 
transitions\cite{ang}. The nature of the curve has significant difference
in the two superconducting states (pure $d_{x^2-y^2}$ and the $d+id$). 
In the $d_{x^{2}-y^{2}}+id_{xy}$ state the specific heat
increases exponentially with temperature, more like the familiar s-wave 
superconductors whereas in the d-wave state the growth is more stiff. This in 
turn indicates that the entropy is higher in pure d-wave state than that in
$d_{x^2-y^2}+id_{xy}$ state. So the low temperature $d+id$ phase, in a way, is 
more ordered than the higher temperature $d_{x^{2}-y^{2}}$ phase.    

The curve for the superfluid density as a function of temperature (Fig. 4)
behaves differently in the two superconducting phases as expected. In the 
$d_{x^{2}-y^{2}}+id_{xy}$ 
phase $\rho_s$ falls exponentially with temperature whereas in $d_{x^{2}-y^{2}}$
phase the descent is according to a power law. At the second transition 
temperature($T_{c2}$), where the 
transition occurs between the two superconducting phases, a sudden 
upturn appears in the $\rho_{s}(T)$ curve which reflects the availability 
of quasiparticle 
excitations due to the disappearance of the $d_{xy}$ state.
If we compare these results with that
of an s-wave superconductor, we observe that the behaviour of $\rho_{s}$ in the 
$d_{x^{2}-y^{2}}+id_{xy}$ state is qualitatively similar to that of the s-wave
state, with a gap all over the FS. Owing to this gap, the 
quasiparticle excitations are not easily accessible at very low temperatures
and keeps the superfluid density almost independent of temperature at 
low temperatures. This exponential behaviour is expected in the 
thermodynamic properties whenever there exists a gap in the excitation 
spectrum. 
  
In conclusion, the thermodynamic properties of the $d_{x^{2}-y^{2}}+id_{xy}$
superconductor are studied with a tight binding model within the 
mean-field theory. Significant differences have been observed in the nature
of the temperature dependence of specific heat and superfluid density between
a pure d-wave state and the $d_{x^{2}-y^{2}}+id_{xy}$ state. The behaviour in
the latter is found to be somewhat similar to that of an s-wave superconductor.
Further experimental observations on the thermodynamics of this state  
will shed light on the microscopic nature of interactions in these 
new class of superconductors. 

\vspace{.5cm} 
\noindent {\bf Acknowledgement} It is a pleasure to thank A. Taraphder for
useful discussions.

\newpage
\center {\Large \bf Figure captions}
\vspace{0.5cm} 
\begin{itemize} 

\item[Fig. 1.] The gap parameters $\Delta_{1}$ and $\Delta_{2}$ (in Kelvin)
versus temperature (in Kelvin). 

\item[Fig. 2.] The quasiparticle energy spectrum ($E_{\bf k}$) (in meV)
along various symmetry directions in the first quadrant of BZ (the gap 
magnitudes have been increased ten times for visualisation). The inset shows
how the symmetry directions are defined in BZ.

\item[Fig. 3] The specific heat versus temperature curve in $d_{x^{2}-y^{2}}+id_{xy}$ and $d_{x^{2}-y^{2}}$ states clearly shows the difference in its
behaviour in these two states. The dotted line shows the normal state
specific heat.
 
\item[Fig. 4] The superfluid density is shown against temperature for 
two phases $d_{x^{2}-y^{2}}+id_{xy}$ and $d_{x^{2}-y^2}$.
 
\end{itemize} 

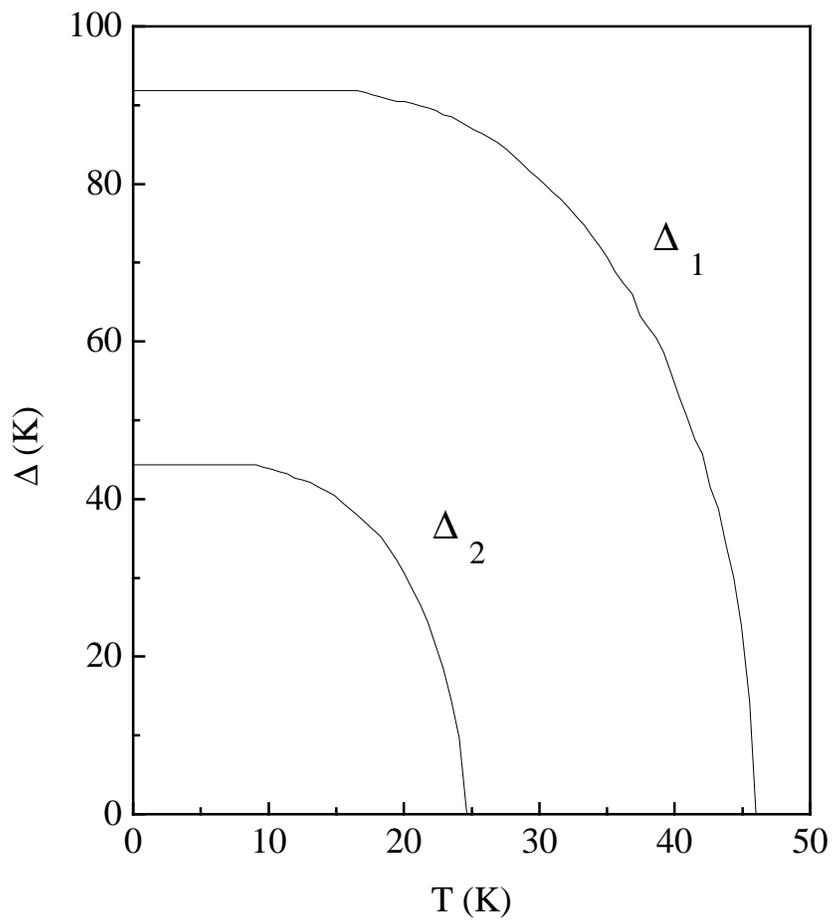
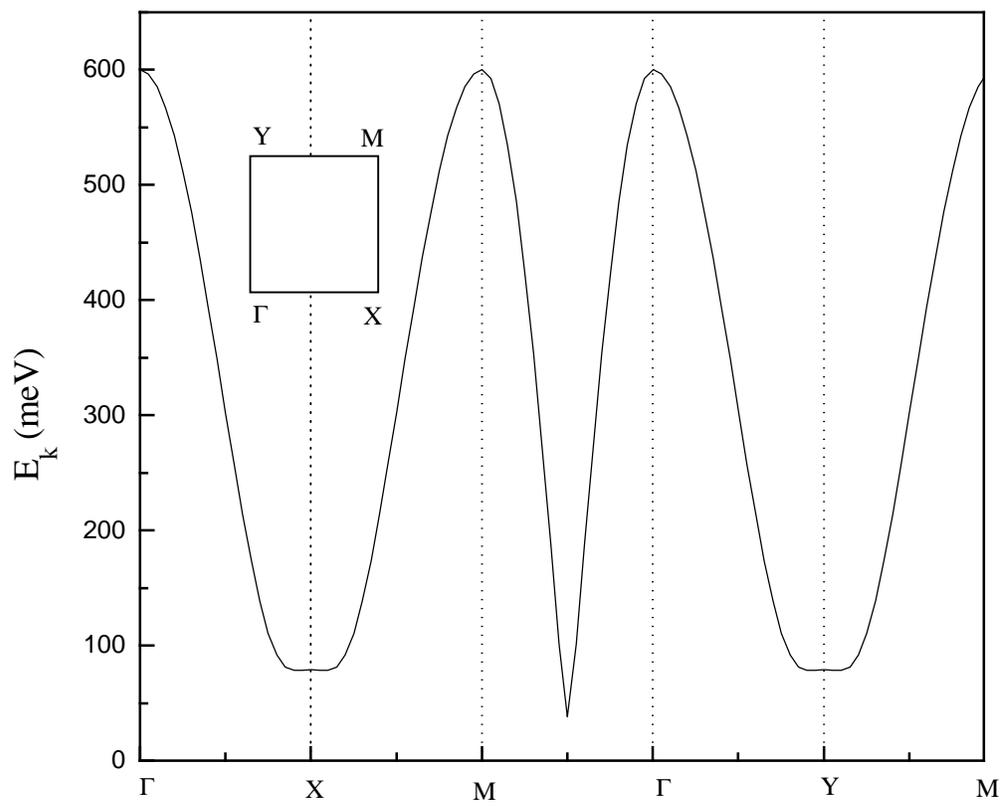
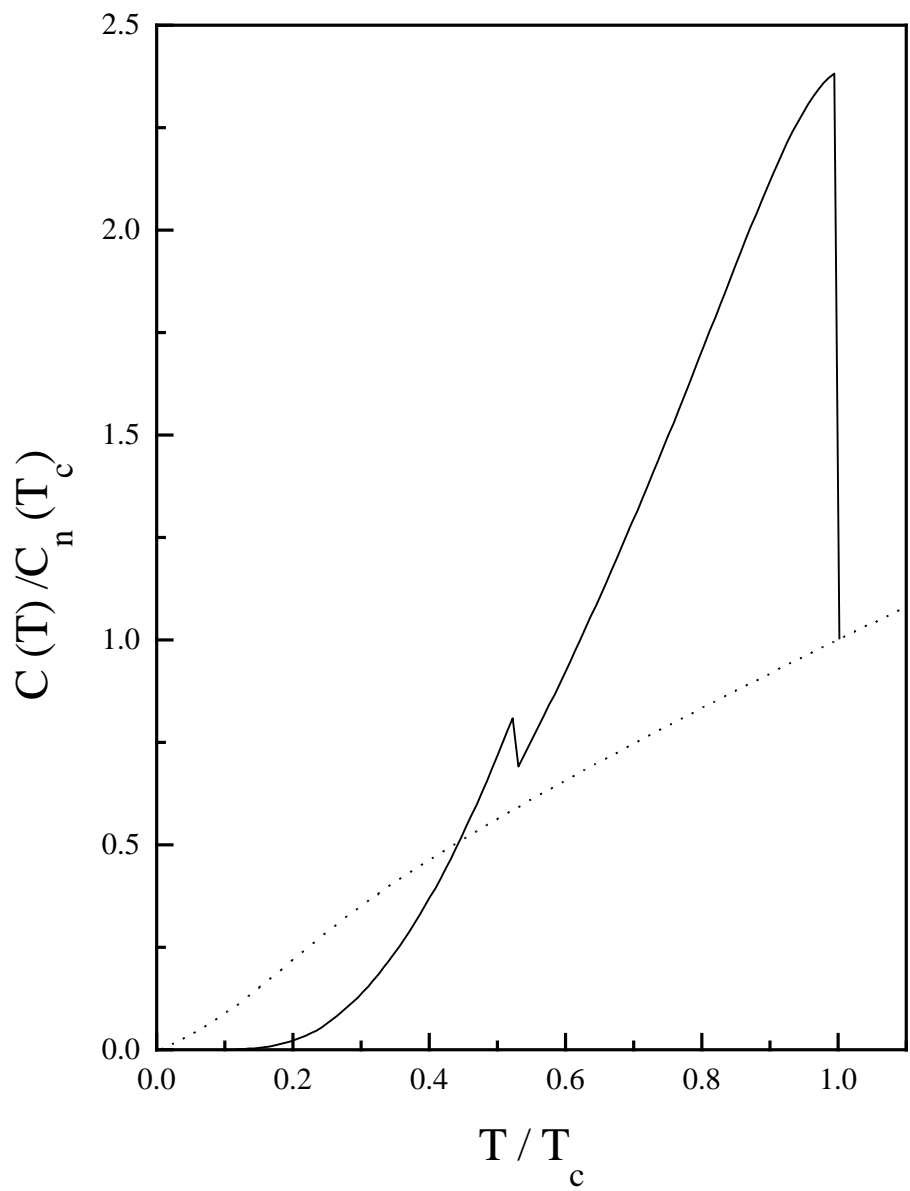
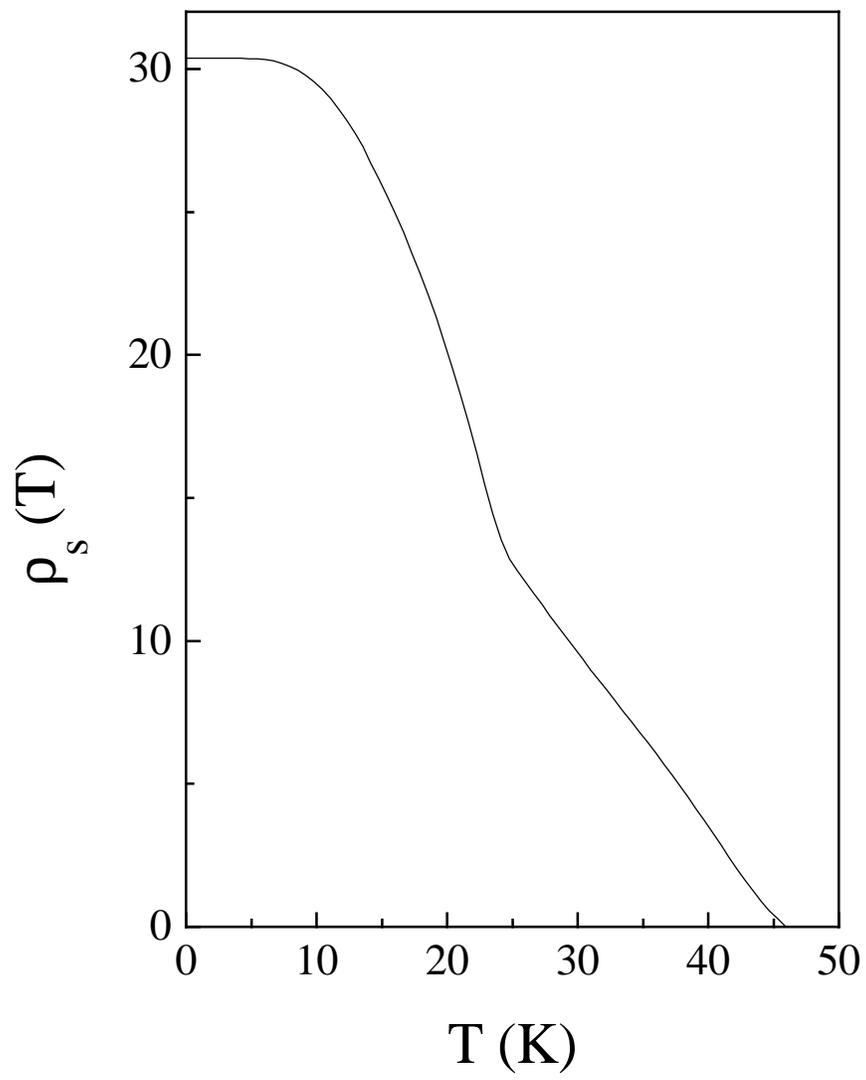
\end{document}